\documentclass[12pt]{iopart} 
\usepackage{graphicx}
\newcommand{\Vec}[1]{\mbox{\boldmath$#1$}}
\begin{document}
\title[Unconventional Pairing in Doped Band Insulators 
on a Honeycomb Lattice]{
Unconventional Pairing in Doped Band Insulators 
on a Honeycomb Lattice : the Role of the Disconnected Fermi Surface and a 
Possible Application to Superconducting $\beta$-MNCl (M=Hf,Zr)}

\author{Kazuhiko Kuroki$^{1,2}$}

\address{$^1$ 
 Department of applied Physics and Chemistry, 
 The University of Electro-Communications, Chofu, Tokyo 182-8585, Japan}
\address{$^2$ 
 JST, TRIP, Chofu, Tokyo 182-8585, Japan}

\ead{kuroki@vivace.e-one.uec.ac.jp}

\begin{abstract}
We investigate the possibility of realizing unconventional superconductivity 
in doped band insulators on the square lattice and the honeycomb lattice, 
where  the latter is found to be a good candidate due to the disconnectivity 
of the Fermi surface. We propose applying the theory 
to the superconductivity in doped layered nitride $\beta$-$M$NCl ($M=$Hf, Zr). 
Finally, we compare two groups of superconductors with 
disconnected Fermi surface, $\beta$-$M$NCl and the iron pnictides, which 
have high $T_c$ despite some faults against superconductivity are present.
\end{abstract}

{\it keywords: } 
$\beta$-MNCl, superconductivity, band insulator, Fermi surface, 
spin fluctuations, pnictides

\section{Introduction}
\begin{figure}[h]
\begin{center}
\includegraphics[width=8cm]{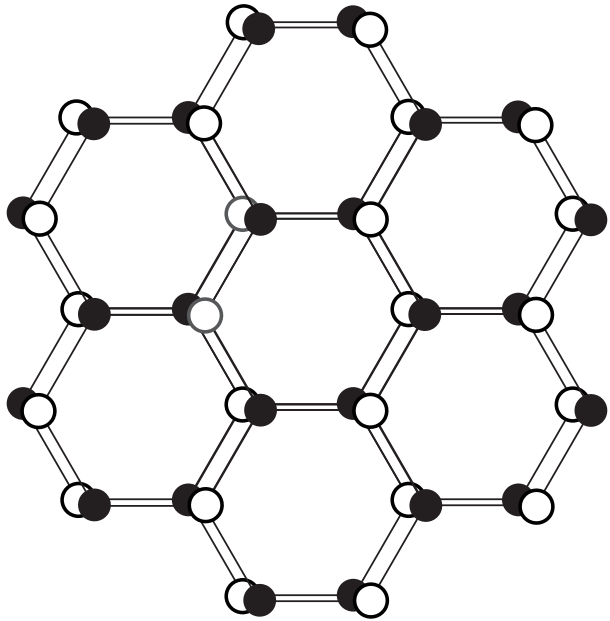}
\caption{\label{fig1}
The bilayer honeycomb lattice is shown. The black and white circles 
represent M(=Hf, Zr) and N, respectively.}
\end{center}
\end{figure}

Layered nitride $\beta$-$M$NCl\cite{Yamanaka} ($M=$Hf, Zr) doped with 
carriers is one of the most interesting group of superconductors.
The mother compound 
$\beta$-$M$NCl is composed of alternate stacking of honeycomb 
$M$N bilayer and Cl bilayer.\cite{Shamoto}
This is a band insulator and becomes a superconductor 
upon doping electrons by Na or Li intercalation.
They exhibit relatively high $T_c$ 
up to $\sim 25$K for $M=$Hf and $\sim 15$K for $M=$Zr.
The bilayer honeycomb lattice structure consisting of $M$ and N 
(Fig.\ref{fig1}) is considered to be playing the 
main role in the occurrence of superconductivity, and the two 
dimensional nature of the superconductivity has been revealed by 
nuclear magnetic resonance\cite{Tou3} 
and muon spin relaxation studies.\cite{Uemura2,Uemura} 
Despite the relatively high $T_c$, 
experimental as well as theoretical studies indicate 
extremely low density of states (DOS)
at the Fermi level.\cite{Tou,Taguchi,Weht} In fact, 
they have the highest $T_c$ 
among materials with the specific heat $\gamma$ as small as 
$\sim$ 1mJ/molK$^2$.
The electron phonon coupling is also 
estimated to be weak,\cite{Tou,Weht,Taguchi2,Heid} and the isotope effect 
is found to be small.\cite{Tou2,Taguchi4}
In the superconducting state, the density of states recover rapidly 
upon increasing the magnetic field,\cite{Taguchi} 
suggesting some kind of anisotropic pairing. 
As for the doping dependence, the DOS at the Fermi level 
stays nearly constant, but, for Li$_x$ZrNCl, 
$T_c$ shows an increase upon lowering the carrier concentration 
until a sudden disappearance of the $T_c$ and a superconductor-insulator
transition is observed.\cite{Taguchi3} On the other hand, in Li$_x$HfNCl, 
$T_c$ stays nearly constant for the whole doping range $x<0.5$.\cite{Takano}
Furthermore, for Li$_x$HfNCl, an intercalation of organic molecules 
tetrahydrofuran (THF) between the layers is found to enhance $T_c$.
\cite{Takano}
While these experiments suggest that some kind of unconventional pairing
mechanism may be at work, 
tunneling spectroscopy experiments on the other hand find an  
$s$-wave like, fully open gap.\cite{Ekino} 
Specific heat measurements also suggest an $s$-wave like gap, 
but again the doping dependence of the magnitude of the gap is unusual.
In the underdoped regime, the gap is large, while the gap becomes small 
as  the doping level is increased, varying from a ``strong coupling'' 
to an ``extremely weak coupling'' superconductor.\cite{Kasahara}

In the present study, we first generally study the possibility of 
unconventional superconductivity by doping band insulators on a 
square or honeycomb lattice.\cite{Bill} We find that the honeycomb lattice is 
a good candidate for realizing such a possibility, where 
the disconnectivity of the Fermi surface plays a role.
Secondly, we consider applying the theory to $\beta$-$M$NCl. The 
two bands sitting closest to the Fermi level in the 
first principles band calculation\cite{Weht} can roughly be 
reproduced by a {\it single} honeycomb lattice model, where the 
above general theory can be applied.
Finally, we compare two groups of superconductors with 
disconnected Fermi surface, $\beta$-$M$NCl and the iron pnictides, which 
have high $T_c$ despite some faults against superconductivity are present.
\begin{figure}[h]
\begin{center}
\includegraphics[width=8cm]{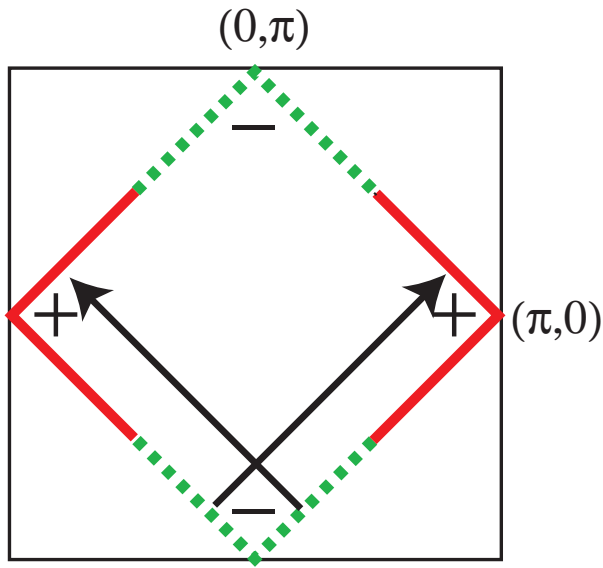}
\caption{\label{fig2} 
An explanation of the $d$-wave superconductivity 
due to spin fluctuations on the square lattice. 
The arrow 
 represents the wave vector $\Vec{Q}$ 
at which the spin fluctuations develop. The solid
 (dashed) lines are the portions of the Fermi surface where the 
 gap has a positive (negative) sign.}
\end{center}
\end{figure}

\section{Square lattice}
Hubbard Model is a model that considers the on-site repulsive 
interaction $U$ on a tight binding model.
Let us start with the Hubbard model on a square lattice, 
where we consider only the nearest neighbor $t$ as the hopping integral.
When the band filling $n$ (=number of electrons/number of sites) 
is near half-filling ($n\sim 1$), 
strong antiferromagnetic spin fluctuations arise, 
and the possibility of $d$-wave superconductivity 
mediated by these spin fluctuations has been 
discussed for the past several decades.
The $d$-wave superconductivity can be understood as follows. 
Generally, superconductivity occurs due to pair scattering mediated 
by the pairing interaction $V(\Vec{q})$.
The gap equation can be written as
\begin{equation}
\Delta(\Vec{k})=-\sum_{\Vec{k'}}\frac{\tanh[E(\Vec{k'})/k_BT]}{2E(\Vec{k'})}
V(\Vec{k}'-\Vec{k})\Delta(\Vec{k}'),
\end{equation}
where $\Delta$ is the gap, $E$ is the quasiparticle dispersion, and 
$T$ is the temperature.
When the spin singlet 
pairing interaction is mediated by spin fluctuations,
$V(\Vec{k})$ is positive and takes large values at 
large $\Vec{Q}$, where the spin fluctuation develops.
So in the case of the square lattice, 
where the spin fluctuations develop near $\Vec{Q}=(\pi,\pi)$, 
we have to change 
the sign of the gap between the wavevectors $\sim (\pi,0)$ and $\sim(0,\pi)$ 
in order to have a finite $\Delta$ as a solution for the 
gap equation, which results in a $d$-wave gap as shown in Fig.\ref{fig2}.
By applying fluctuation exchange (FLEX) approximation to this system, 
which is a kind of self-consistent perturbation theory that collects 
random phase approximation type diagrams, we can obtain the Green's 
function and the spin susceptibility.\cite{Bickers} 
These can be plugged into the Eliashberg equation, whose solution 
gives a $d$-wave superconductivity with a $T_c$ of 
the order of $0.01t$, where $t$ is the 
nearest neighbor hopping integral (if $t\sim 1$eV, $T_c$ is of the 
order of 100K). 

\begin{figure}[h]
\begin{center}
\includegraphics[width=35pc]{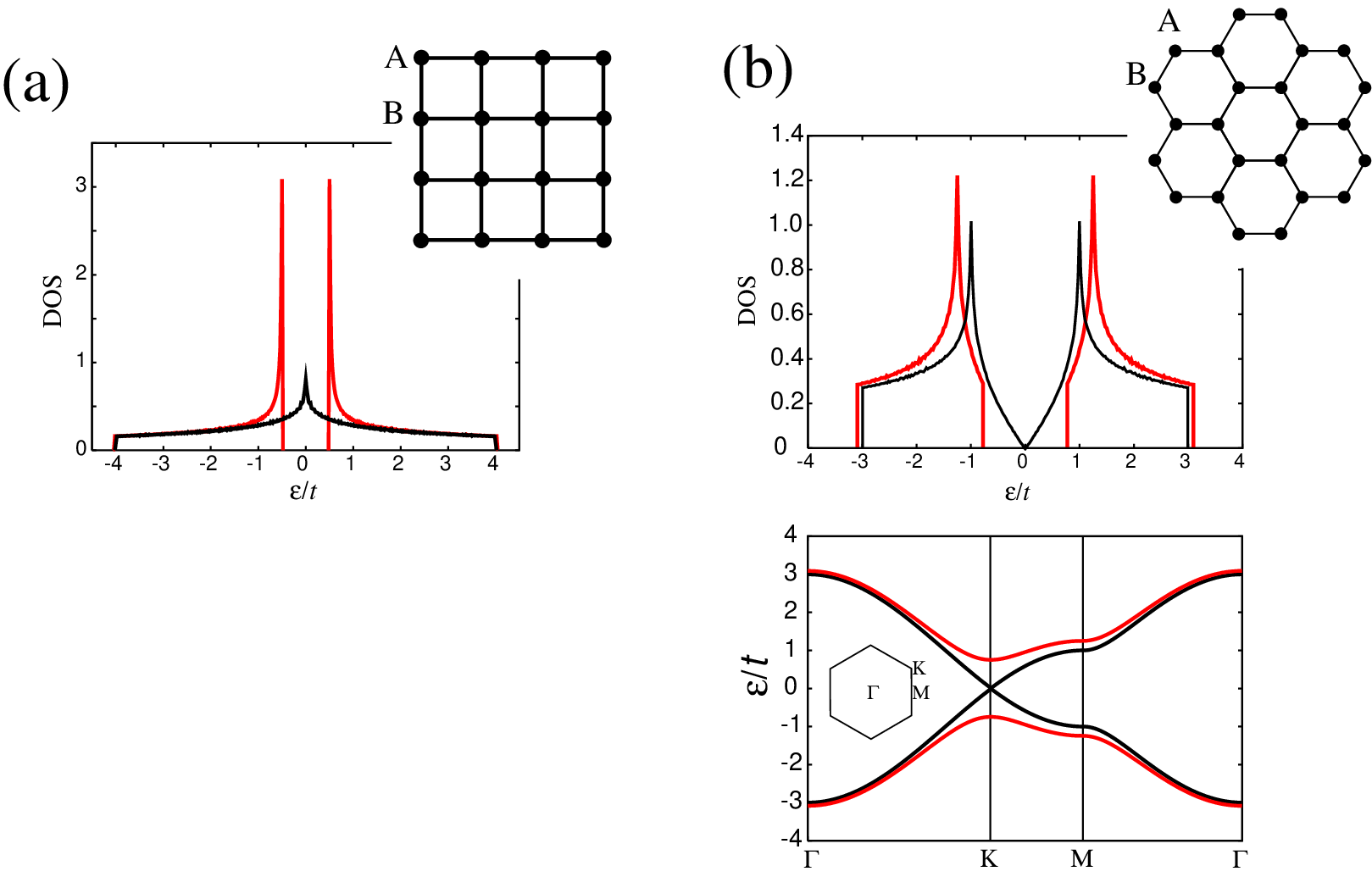}
\caption{\label{fig3}
(a) The density of states for the square lattice 
with $\Delta=0$ (black) and $\Delta=t$ (red).(b) The density of 
states (upper) and the band dispersion (lower) 
for the honeycomb lattice with $\Delta=0$ (black) and $\Delta=1.5t$. 
The hexagonal Brillouin zone of the honeycomb lattice is shown in the 
inset of the lower panel.}
\end{center}
\end{figure}

\begin{figure}[h]
\begin{center}
\includegraphics[width=15pc]{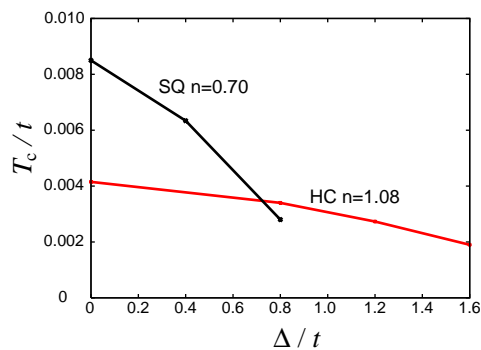}
\caption{\label{fig4} $T_c$ as functions of the level offset $\Delta$ 
obtained by FLEX+Eliashberg equation for 
the square lattice with $U=6t$, $n=0.7$ (black) or for the honeycomb 
lattice with $U=6t$ and $n=1.08$ (red).}
\end{center}
\end{figure}

Now, the square lattice (with only the nearest neighbor hopping) 
is a bipartite lattice which can be separated into A and B sublattices. 
Let us see what happens to the superconductivity 
if we introduce a level offset $\Delta$ between A and B lattices.
The introduction of $\Delta$ opens up a gap at the center of the band, so 
this amounts to investigating the possibility of 
unconventional superconductivity by doping carriers in 
band insulators (Fig.\ref{fig3}(a)). 
As shown in Fig.\ref{fig4}, 
we find that the introduction of $\Delta$ rapidly suppresses
superconductivity. Thus, in the case of the square lattice, 
chances for realizing unconventional superconductivity in the above sense 
are small.

\begin{figure}[h]
\begin{center}
\includegraphics[width=35pc]{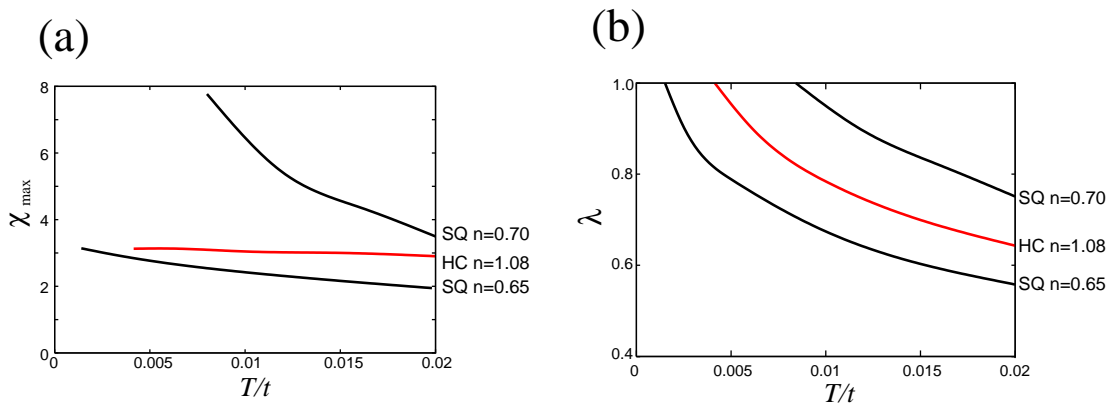}
\caption{\label{fig5}(a) The maximum value of the spin susceptibility 
as functions of temperature obtained by FLEX 
for the square lattice with $n=0.7$ or $n=0.65$ and 
for the honeycomb lattice for $n=1.08$. $U=6t$ in all cases. 
HC and SQ stand for the honeycomb and the square lattices, 
respectively.
(b) The eigenvalue of the linearized Eliashberg equation with 
the same parameter values as in (a). The 
temperature at which $\lambda=1$ is the $T_c$.}
\end{center}
\end{figure}

\begin{figure}[h]
\begin{center}
\includegraphics[width=23pc]{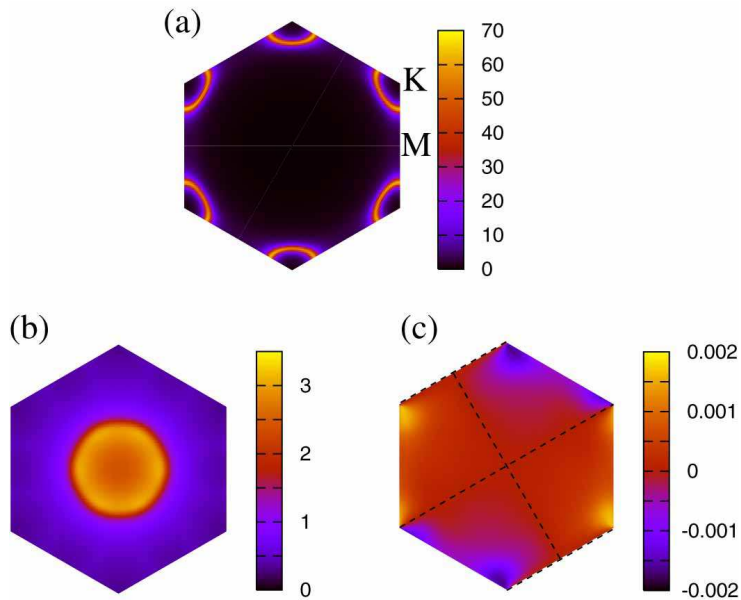}
\caption{\label{fig6}The contour plots of the FLEX result 
at the lowest Matsubara frequency for 
the honeycomb lattice  in the hexagonal Brillouin zone
with $U=6t$, $n=1.08$, and $T=0.01t$ (a) The 
Green's function of the upper band squared,  
(b) the spin susceptibility, (c) the superconducting 
gap function.}
\end{center}
\end{figure}

\begin{figure}[h]
\begin{center}
\includegraphics[width=23pc]{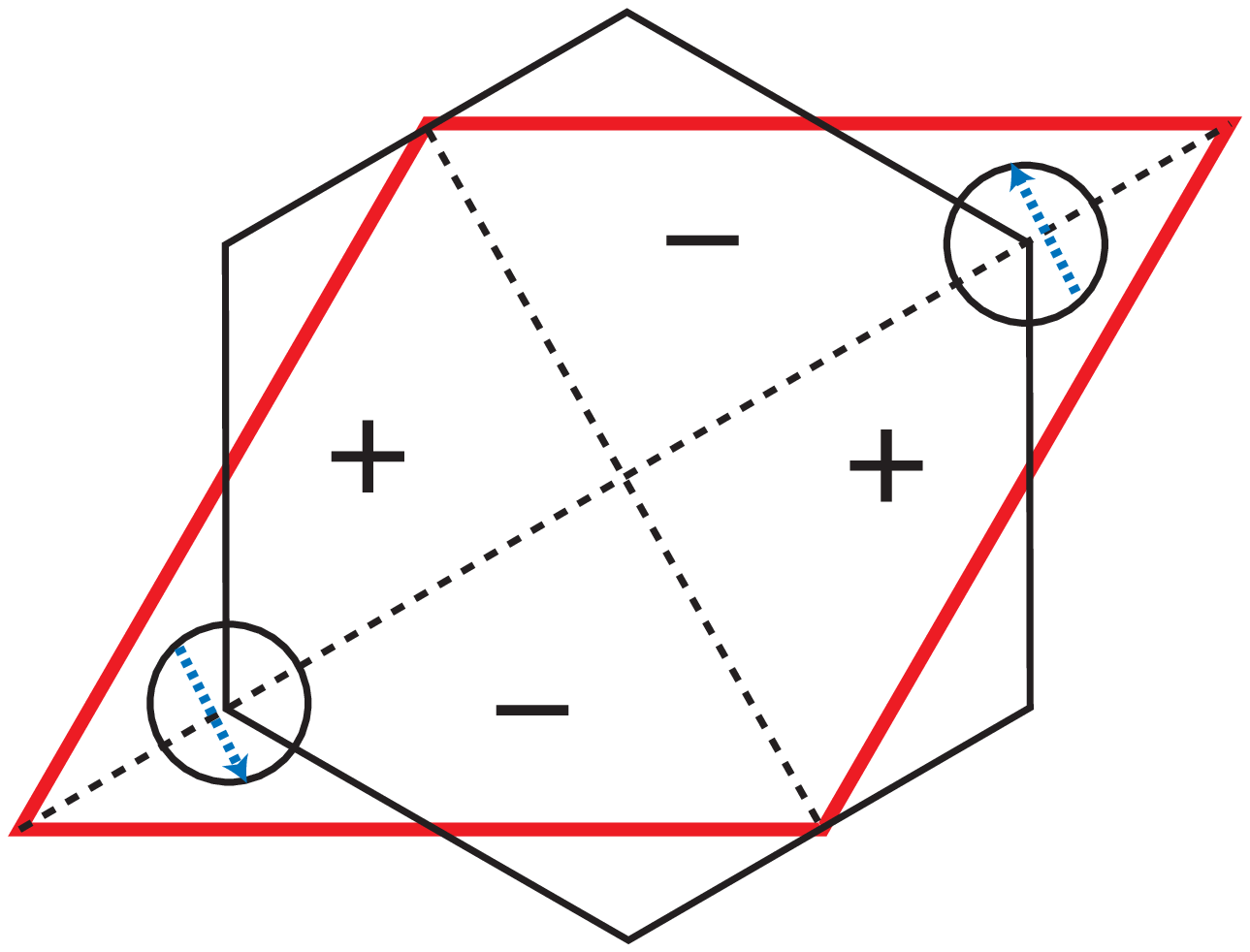}
\caption{\label{fig7} The Fermi surface (the two circles) 
and the sign of the gap function are 
schematically shown in the extended zone scheme. The dashed arrows 
represent the wave vectors at which the spin fluctuations develop.}
\end{center}
\end{figure}

\section{Honeycomb lattice}
Now, let us compare the above result for the square lattice with those for 
another two dimensional bipartite lattice, namely, the honeycomb lattice.
The honeycomb lattice is a system with two sites in a unit cell, 
in which the two bands make point contact at K and K' points of the 
Brillouin zone,  
resulting in a zero gap density of states (Fig.\ref{fig3}(b)).
We show in Fig.\ref{fig5}(a) the maximum value of the spin 
susceptibility as a function of temperature 
for the band filling of $n=1.08$, 
which corresponds to a small amount of electron doping. 
Surprisingly, we find that the spin susceptibility 
is nearly independent of $T$, 
which is in sharp contrast with the 
case of the square lattice. For example, for the 
square lattice with $n=0.7$, which is already substantially 
away from half-filling, 
we still have a strong enhancement of the spin susceptibility upon 
lowering the temperature. 
With further hole doping to $n=0.65$, the spin susceptibility 
is suppressed, but even in that case, there is a 
moderate increase of the spin susceptibility upon lowering the temperature.
In Fig.\ref{fig5}(b), 
we show the eigenvalue $\lambda$ 
of the linearized Eliashberg equation as functions of 
temperature. $T_c$ is the temperature where $\lambda$ reaches unity. 
The density of states at 
$E_F$ is nearly the same for the square lattice with $n=0.65$ and the 
honeycomb lattice with $n=1.08$, and 
also the spin susceptibility has similar values at low temperature, 
but still, the honeycomb lattice has larger $\lambda$ 
and higher $T_c$. Thus, the Hubbard model on the 
honeycomb lattice has relatively high $T_c$ despite the 
low density of states and the weak and temperature 
independent spin fluctuations.

Fig.\ref{fig6} shows the contour plot of the Green's function squared, 
whose ridge corresponds to the Fermi surface. 
We see here two disconnected pieces of the Fermi surface. 
The spin susceptibility is maximized at wave vectors that bridge the 
opposite sides of each pieces of the Fermi surface. 
As can be seen more clearly in Fig.\ref{fig7}, 
the gap has a $d$-wave form, where the gap changes sign across the 
wave vector at which the spin susceptibility is maximized.
Here, note that one of the 
nodes of the gap do not intersect the Fermi surface 
because of its disconnectivity,
which may be one reason why superconductivity 
is favored despite the low density of states and the weak spin fluctuations.
By symmetry, there are two degenerate $d$-wave gaps 
(say, $d_{xy}$ and $d_{x^2-y^2}$, or any two linearly independent  
combinations),  and the most probable form of the gap 
below $T_c$ is the form $d+id$, where the two $d$-wave gaps mix with a 
phase shift of $\pi/2$. Since the two $d$-wave gaps have nodal lines at 
different positions, this kind of mixture leads to a gap that has a 
finite absolute value on the entire Fermi surface. An important point 
is that if such a state is realized, the time reversal symmetry should 
be broken.

Now we introduce the level offset between A and B sites 
as we did for the square lattice. In this case also, 
a band gap opens in the center (Fig.\ref{fig3}(b)), 
so we once again investigate the possibility of superconductivity by 
doping band insulators. In this case, 
we find that superconductivity is relatively robust against the introduction 
of $\Delta$. This may be because the density of states is 
already low in the original honeycomb lattice, 
so that the introduction of $\Delta$ 
does not affect superconductivity so much.
\begin{figure}[h]
\begin{center}
\includegraphics[width=23pc]{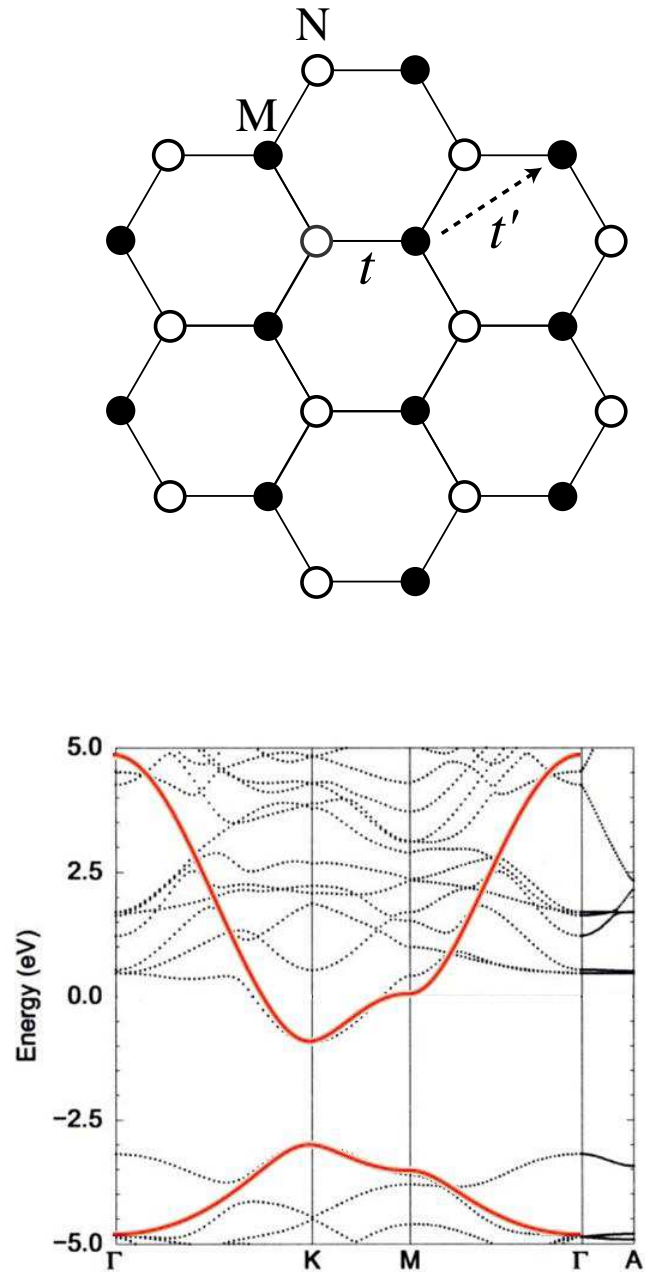}
\caption{\label{fig8}Upper panel: the tightbinding model considered for
 $\beta$-MNCl. Lower panel: the first principles band calculation taken 
from ref.\cite{Weht} and the band dispersion of the tightbinding model.}
\end{center}
\end{figure}

\begin{figure}[h]
\begin{center}
\includegraphics[width=23pc]{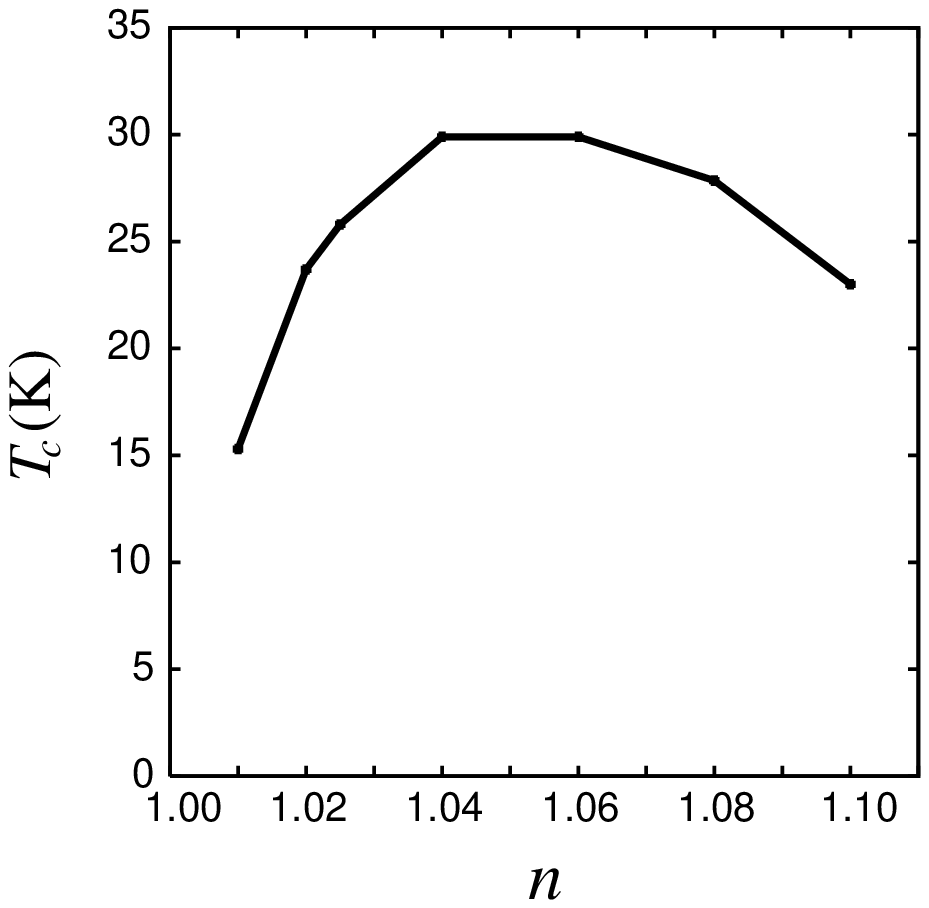}
\caption{\label{fig9} $T_c$ plotted as a function of the band filling for the 
model shown in Fig.\ref{fig8}. }
\end{center}
\end{figure}

\section{Application to $\beta$-MNCl}

We now consider applying the above theory to $\beta$-MNCl.
Although $\beta$-MNCl has a bilayer honeycomb lattice structure,  
we find that the two bands 
closest to $E_F$ obtained in the first principles calculation
\cite{Weht,Heid,Hase,Oguchi} 
can roughly be reproduced by a {\it single} layer honeycomb lattice 
model consisting of alternating ''M'' and ''N'' 
orbitals with a level offset 
as shown in Fig.\ref{fig8}. Here we take $t=1.2$eV, $\Delta/t=2.7$, 
and $t'/t=0.35$.
If we consider on-site repulsive interaction $U=6t$ 
on both M and N orbitals,\cite{comment} the model is similar to the one 
studied in section 2, except that  distant hopping integrals $t'$ 
have to be considered so as to reproduce the first principles band 
structure. 
Consequently, within the FLEX+Eliashberg equation approach, 
we obtain relatively high $T_c$ of around 30K as shown in Fig.\ref{fig9}.

Now we compare the present scenario with the experiments for $\beta$-MNCl.
As mentioned in the Introduction, relatively high $T_c$ is obtained despite 
the extremely low density of states and the weak 
electron phonon coupling,\cite{Tou,Taguchi,Taguchi2,Weht,Heid} 
which can be explained within the present theory.
The isotope effect is small,\cite{Tou2,Taguchi4} 
which again seems consistent since the 
the present pairing mechanism is purely electronic.
The cointercalation of THF molecules enhances $T_c$,\cite{Takano} 
and this also seems 
to be understandable  within this kind of spin fluctuation 
mediated pairing, where the quasi two dimensionality is favored 
as discussed generally in refs.\cite{Arita,Monthoux}.
As for the pairing symmetry, 
a fully open gap is observed in the experiments,\cite{Ekino} and 
this is consistent with the present scenario provided that the $d+id$ state 
is realized. It is hence interesting to investigate experimentally 
the possibility of time reversal symmetry breaking in the 
superconducting state of this material.
The rapid recovery of the specific heat by applying 
the magnetic field,\cite{Taguchi} 
and also the unusual doping dependence of both the $T_c$ and the 
magnitude of the gap\cite{Taguchi3,Takano,Kasahara} 
remain as an interesting future problem.

\section{Superconductivity in systems with disconnected Fermi surface}

\begin{figure}[h]
\begin{center}
\includegraphics[width=6cm]{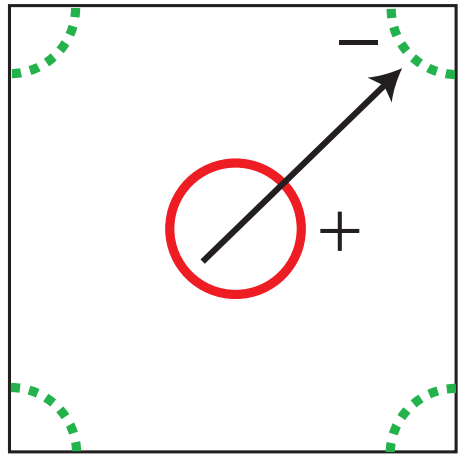}
\caption{\label{fig10} An example of fully open sign reversing gap  
on a  disconnected Fermi surface is shown. The arrow shows the nesting 
vector at which the spin fluctuations develop.}
\end{center}
\end{figure}

Finally, let us go back to the $d$-wave superconductivity on 
the square lattice. As mentioned, 
$T_c$ obtained by FLEX is the order of $0.01t$, where $t$ 
is the hopping integral.
This can correspond to a high $T_c$ in actual materials 
because $t$ can be of the order of electron volt, 
but still this is low compared to the original energy scale $t$.
There are several reasons for the ``low'' $T_c$, and one of them is that 
we have to have nodes of the gap {\it intersecting} the Fermi surface because 
sign change of the gap is required in the case of spin 
fluctuation mediated pairing.
In this context, we proposed some time ago 
that if there are disconnected pieces of Fermi surface 
nested to some extent, 
we can change the sign of the gap between the disconnected peaces 
without the nodes intersecting the Fermi surface, and this can 
result in a high $T_c$ superconductivity.\cite{KA} An example of 
such a Fermi surface is shown in Fig.\ref{fig10}.
Possibilities of the disconnected Fermi surface playing important 
role in the occurrence of superconductivity or the determination of the 
pairing symmetry have been discussed for a cobaltate 
Na$_x$CoO$_2$\cite{KKcobalt} and 
an organic superconductor (TMTSF)$_2$X.\cite{KKtmtsf}

\begin{figure}[h]
\begin{center}
\includegraphics[width=35pc]{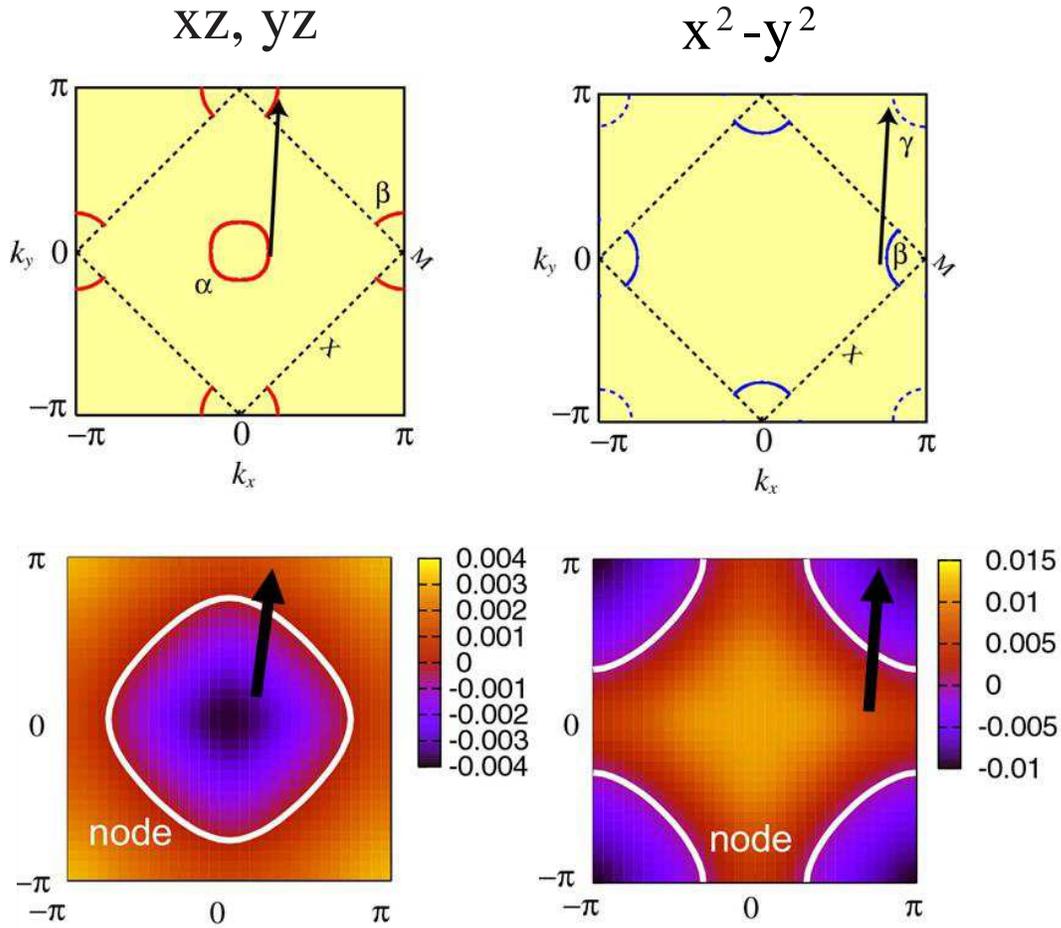}
\caption{\label{fig11} Upper panels : the portions of the Fermi surface 
of iron pnictides with strong $d_{xz}$, $d_{yz}$ (left) or $d_{x^2-y^2}$ 
(right) orbital 
character. The dashed curve around $(\pi,\pi)$ (at the corners of the 
Brillouin zone) is the portion where a 
flat portion of the band lies 
very close to the Fermi level, although it does not 
actually produce a Fermi surface. 
The arrows show the nesting vector of the Fermi surface. 
Lower panels: the gap function for the 
$d_{xz}$, $d_{yz}$ (left) or $d_{x^2-y^2}$ (right) orbitals. }
\end{center}
\end{figure}

Quite recently, superconductivity has been found in 
iron-based pnictides by Hosono and coworkers.\cite{Hosono} 
The highest $T_c$ of this series of materials have reached 55K.\cite{Ren}
Band calculations show that there are several 
disconnected pieces of the Fermi surface in this material,\cite{Singh} 
where spin fluctuations can arise due to the nesting between them.\cite{Mazin} 
According to our Eliashberg theory calculation that takes into 
account such kind of spin fluctuations,\cite{KKpnic} 
the gap changes sign across the nesting vector of the Fermi surface, 
and the magnitude of the gap 
is especially large on the portion of 
the Fermi surface where the $d_{x^2-y^2}$ orbital character is strong 
as shown in Fig.\ref{fig11}. 
The nearest neighbor hopping of this orbital is about 
0.15eV, which is 
quite small, and the experimentally observed maximum $T_c=55$K  
corresponds to about $0.04t$ which is higher than that can 
be reached in the single band square lattice.

So the two groups of materials seem to exhibit high $T_c$ despite 
some kind of faults against superconductivity are present:  
for $\beta$-MNCl, relatively high $T_c$ is obtained despite the extremely low 
DOS and the weak spin fluctuations (and electron-phonon coupling), while 
for the pnictides, high $T_c$ is obtained despite the low energy scale of the 
main band, and also competing spin and charge fluctuations due to the 
multiplicity of the orbitals. 
The disconnectivity of the Fermi surface 
may be one good reason why these faults are overcome.
In the future, there may a possibility that 
we can get higher $T_c$ by realizing disconnected 
Fermi surface on more ideal situations. 

The author acknowledges Y. Taguchi, Y. Iwasa, Y. Kasahara, T. Ekino, 
and H. Aoki for fruitful discussions.
Numerical calculation has been done at the computer center, ISSP, University 
of Tokyo. This study has been supported by Grants-in Aid from MEXT of Japan 
and JSPS.


\begin{thebibliography}{9}
\bibitem{comment} Generally, the on-site repulsion on the 
M site and N site can be different, but here we assume them to be 
equal for simplicity.
\end{thebibliography}
\end{document}